\documentclass[twocolumn,aps,pra]{revtex4}
\begin{document}
\title{Heisenberg chains cannot mirror a state.}
\author{Marcin Wie\'sniak}
\affiliation{Centre for Quantum Technologies, National University of Singapore, 3 Science Drive 2, Singapore 117543, Singapore}
\begin{abstract}
In the future, faithful exchange of quantum information can become a key part of many computational algorithms. Some authors suggest using chains of mutually coupled spins as channels for quantum communication. One can divide these proposals into the groups of assisted protocols, which require some additional action from the users, and natural ones, based on the concept of state mirroring. We show that state mirroring is fundamentally not the feature of chains of spins-$\frac{1}{2}$ coupled by the Heisenberg interaction, but without local magnetic fields. This fact has certain consequences in terms of the natural state transfer.
\end{abstract}
\maketitle
Transferring information is one of the most important parts of classical computing. It is also the simplest. Classical wires are just pieces of metal, in which the electromagnetic field can freely propagate.

Quantum Mechanics also offers interesting computation algorithms \cite{jozsa,grover,shor}, which can be implemented in various physical systems. In measurement-based techniques, such as the cluster state computation \cite{cluster}, there is no need to exchange quantum information, but this is at the expense of maintaining a large entangled state. One may say that in this computational process the information was initially exchanged between all the qubits in the lattice, later to be only processed locally.

In proposals involving also unitary transformations (gates), the exchange of quantum information is a challenge. One can use the teleportation \cite{teleportation}, again dependent on entangled states. They are often considered an additional resource and could be not easy to produce. The exchange of photons between more distant sites of the lattice is not a good solution as there is no faithful and controllable atom-photon interface \cite{alicki}. The other, often discussed, idea is sending the state through a quantum wire, a chain (or more generally, a graph) of mutually coupled qubits.

Before we prove the title statement, let us briefly discuss some state transfer proposals, together with their possible disadvantages. 
The original idea of quantum communication through a system of interacting spins-$\frac{1}{2}$ was introduced by Bose \cite{boseprl91207901}. He has considered an arbitrary Heisenberg lattice described by the following Hamiltonian:

\begin{equation}
H_1=\sum_{i,j=1}^NJ_{ij}\vec{\sigma}^{[i]}\cdot\vec{\sigma}^{[j]}+\sum_{i=1}^NB_i\sigma_z^{[i]}.
\end{equation}
$\vec{\sigma}^{[i]}$ denotes the usual Pauli matrix vector acting on a qubit labeled $i$, and $\sigma_z^{[x]}$ is the $z$ component thereof. $J_{ij}$ denotes a coupling constant between two spins, and the local magnetic field is $B_i$.

A qubit can be encoded as follows: for the logical $|0\rangle$ we take the fully magnetized state, $|00...0\rangle$, while logical $|1\rangle$ is translated into the flip of the first spin (the one closest to the sender), $|10...0\rangle$. Due to the free evolution, the excitation $|1\rangle$ wanders over the whole system. The idea of the protocol is simply to wait until the time, at which the average fidelity of the receiver's qubit to the input state, $F=\int\langle \psi_{in}|\rho_{out}|\psi_{in}\rangle d\mu(|\psi_{in}\rangle)$ ($\int d\mu(|\psi_{in}\rangle)=1$) has a peak. 

Bose discusses an example of the uniform Heisenberg chain ($J_{ij}=\frac{J}{2}\delta_{j,i+1}, B_i=B$). For the number of spins $N\approx 80$ and the maximum time of waiting $t_{\max}=4000/J$, he observes $F$ to be hardly above the classical limit, equal to $2/3$. This shows the main problem in quantum state transfer. If energies are not mutually rational, the excitation is irreversibly spread over the wire. At some instances, the eigenstates interfere constructively enough to partially extract encoded information from the last spin, but in general, there is always some part of it left in the chain and hence lost to the receiver.

In fact, the periodic evolution is a feature of very few lattices. The Reader should keep in mind that a complex quantum protocol could involve a number of transfer routines. On top of that, the environment interacts with the system causing decoherence \cite{danielbose,chiny,kay1,exp3,thermal}. The cumulative effect of these losses could cancel the benefit of quantum computing.

Bose has noticed, however, that even with a low fidelity of the transfer, one can constitute entanglement between the sender's and the receiver's spins. This entanglement could be accumulated over many runs, distilled \cite{distill}, and used to teleport the message.

Subsequent solutions \cite{christ,niko1,niko2,Li,Shi,christ1,danielbose,danielbose1,bgb,dago,paternostro} were aimed to minimize the loss of information. An  example is ``the valve scheme'', which was introduced in \cite{bgb}. This time, apart from the wire, one has an additional qubit (``the bucket''). There is also a gate (``the valve'') between the bucket and the last spin of the chain is occasionally switched on by the receiver. This gate is time-dependent partial SWAP operation. Similarly as in the first scheme, logical $|0\rangle$ and $|1\rangle$ are represented by the magnetized state, and an excitation in the chain, respectively. The partial SWAP is designed in such a way, that it transfers the excitation only from the chain to the bucket. One needs to perform this operation many times to gather the complete information (in the asymptotic limit).

The secret behind the action of the valve is that it is being adjusted to the free evolution. In the first iteration, it simply acts as a full SWAP gate, as the bucket was initialized in $|0\rangle$. Then we let the chain to freely evolve for some time, until we again switch on the valve. This time it needs to take into account the present state of the bucket. To use this protocol, one needs to precisely simulate the free dynamics. The good news is, it is enough to do the simulation only once and devise a circuit, which would ``replay'' it for every transfer. Still, the protocol requires a huge amount of computational power, as for such a simple process. 

An example of a lossless protocol was proposed by Burgarth and Bose in \cite{danielbose}. Instead of using one chain of spins, their dual-rail protocol involves two independent, arbitrary, but identical such systems (the scheme was later generalized for two non-identical systems \cite{danielbose1}). The input is encoded by injecting the excitation into one or the other chain. At the output side, the magnetization of the two last spins, $\sigma_z^{[N,1]}+\sigma_z^{[N,2]}$, is measured at some time determined numerically. If the result is 0, the state of the two receiver's qubits is the same as initially created by the sender. This measurement can be repeated until success. 

An important feature of the dual rail protocol is that whenever the transfer is conclusive, it is also faithful. This is, needless to say, in the idealized case. The superposition can be much deformed if the two chains are not identical and the measurement is conducted after rather a long time. This protocol can also turn out inefficient if one includes decoherence. Two chains are a much bigger system than a single one, which therefore has much more ways to interact with the environment. 

The above schemes, among others (e.g., \cite{dago,paternostro}), are examples of assisted state transfer protocols. Their success comes at the expense of an additional action taken by the user. This is as opposed to natural schemes, in which the only conducted operations are encoding and decoding. These proposals rely on mirroring chains \cite{mirror}. A mirroring chain is symmetric with respect to its middle. The parity (symmetry or antisymmetry) of its Hamiltonian eigenstates matches the parity of the respective energy. Up to irrelevant common additive and multiplicative constants, the even states are expected to have even integer energies, and odd states are associated to odd eigenvalues. This is known as the spectrum parity-matching condition (SPMC) \cite{Shi}. As a concequence, at half of the period, the relative phase between the odd and the even components of the state is changed by $\pi$. Whatever was initialized at one end of the chain, can be now found at the opposite. An example of a system with this property was found by Christandl {\em et al.} \cite{christ}, and independently by Nikolopoulos, Petrosyan, and Lambropoulos \cite{niko1,niko2}. It is an $xx$ chain, the Hamiltonian of which is given by
\begin{equation}
\label{h2}
H_2=\sum_{i=1}^{N-1}J_{i,i+1}(\sigma_x^{[i]}\sigma_x^{[i+1]}+\sigma_y^{[i]}\sigma_y^{[i+1]}).
\end{equation}
For the choice of coupling constants $J_{i,i+1}=\sqrt{i(N-i)}$, the fully magnetized state has energy 0, while the one excitation subspace is built of states of energies $-\frac{N-1}2,-\frac{N-3}2,...,\frac{N-3}2,\frac{N-1}2$.

Different physical systems realize different couplings. The $xx$ interaction is typical for opto-atomic implementations \cite{cavities}, but also more general interactions can be observed in those systems \cite{nadeslane}. The Heisenberg interaction is specific for some solid state technologies, e.g., quantum dots \cite{dots1,dots2,dots3}.

The Hamiltonian with the above coupling constants fails to be a state mirroring system if we replace the $xx$ interaction between qubits with the latter type. In what remains, we are going to consider the possibility of a natural state transfer between sites 1 and $N$ in a chain of $N$ spin-$\frac{1}2$ with the isotropic interaction between the nearest neighbors. As we argue at the end, this particular process relies on the general feature of state mirroring in the one excitation subspace, and the possibility thereof will be of our interest.

The problem is trivial if one allows local magnetic fields, specifically chosen for each qubit. This would make it possible to cancel the diagonal terms of the Hamiltonian (projected onto the one excitation subspace) and bring it to the form of Eq. (\ref{h2}). Such a precise adjustment of the magnetic field could be, nevertheless, challenging for a system of physical length of a few micrometers. We are rather interested in the case of a uniform magnetic field acting over the whole wire. The magnitude of this field is here chosen to be 0.

As shown by Katsura \cite{katsura}, the Hamiltonian of a linear system with $xx$ interaction between nearest neighbors can be expressed in terms of a finite field of noninteracting fermions. This allows the chain presented in \cite{christ,niko1,niko2} to mirror all states, not only those in particular subspaces, provided a proper choice of the magnitude of the uniform magnetic field. $\sigma_z^{[i]}\sigma_z^{[i+1]}$ represents the interaction between two fermionic modes. After the Jordan-Wigner transformation, which constitutes the canonical anticommutation relations,
\begin{eqnarray}
a^{[k]}=&\frac{1}{2}\left(\prod_{j<k}\sigma_z^{[j]}\right)\left(\sigma_x^{[k]}+i\sigma_y^{[k]}\right)&,\\
a^{\dagger[k]}=&\frac{1}{2}\left(\prod_{j<k}\sigma_z^{[j]}\right)\left(\sigma_x^{[k]}+i\sigma_y^{[k]}\right)&,
\end{eqnarray}
the $zz$ term takes the form of
\begin{equation}
\label{zzJW}
\sigma_z^{[i]}\sigma_z^{[i+1]}=(2a^{\dagger[i]}a^{[i]}-1)(2a^{\dagger[i+1]}a^{[i+1]}-1).
\end{equation}

It is important to notice that the Heisenberg Hamiltonian, as well as any $xx$ or $xxz$ model, commutes with the total magnetization in the $z$ direction, $\sum_{i=1}^N\sigma_z^{[i]}$. Hence a fixed $z$ magnetization of the initial state (or its component) will be preserved throughout the whole free evolution.

It suffices to consider the property of mirroring not in general, but only in certain subspaces--including the fully magnetized state (the fermionic vacuum) and the one excitation space. These subspaces are involved in state transfer protocols.  For the sake of our analysis, we can drop the first term, as we aim to work with one fermion only, and the irrelevant constant. The one excitation part of the Hamiltonian (the projection of $H$ onto the subspace of one fermion states) has the three-diagonal form and reads 
\begin{widetext}
\begin{equation}
H_{oe}=-\frac{1}{2} \left(\begin{array}{cccccc}-J_{1(N-1)}&J_{1(N-1)}&0&...&0&0\\ J_{1(N-1)}&-J_{1(N-1)}-J_{2(N-2)}&J_{2(N-2)}&...&0&0\\ 0&J_{2(N-2)}&-J_{2(N-2)}-J_{3(N-3)}&...&0&0\\ ...&...&... & &...&...\\ 0&0&0&...&-J_{1(N-1)}-J_{2(N-2)}&J_{1(N-1)}\\0&0&0&...&J_{1(N-1)}&-J_{1(N-1)}	\end{array}\right).
\end{equation}
\end{widetext}
$J_{i(N-i)}$ is the coupling constant and the structure of the subscript stresses the spatial symmetry of the chain.

The first step is to separate the odd and even sectors with a unitary operation
\begin{equation}
U=\frac{1}{\sqrt{2}}\left(\begin{array}{ccccc} 1&0&...&0&1\\ 0&1&...&1&0\\ ...&...& &...&...\\ 0&1&...&-1&0\\ 1&0&...&0&-1\end{array}\right)
\end{equation} 
for $N$ even and
\begin{equation}
U=\frac{1}{\sqrt{2}}\left(\begin{array}{ccccccc}1&0&...&...&...&0&1\\ 0&1&...&...&...&1&0\\...&...&...&...&...&...&...\\ ...&...&0&\sqrt{2}&0&...&...\\...&...&...&...&...&...&...\\0&1&...&...&...&-1&0\\1&0&...&...&...&0&-1\\ \end{array}\right)
\end{equation}
for $N$ odd. The one excitation part of the Hamiltonian is transformed into two three-diagonal matrices, which we will call $H_e$ and $H_o$:
\begin{equation}
UH_{oe}U^\dagger=\left(\begin{array}{cc}H_e&0\\ 0&H_o\end{array}\right).
\end{equation}

If the Hamiltonian expresses a periodic evolution, it can always be rescaled so that all eigenvalues are integers. Having done that, we are looking for such values of $J_{i(N-i)}$s, which satisfy the SPMC that all the eigenvalues of $H_e$ are even and all the eigenvalues of $H_o$ are odd. The first hint is obtained from the comparison of the traces. 
\begin{equation}
\text{Tr}(H_e)-\text{Tr}(H_o)=J_{i_0(N-i_0)},
\end{equation}
with $i_0=N/2$ or $(N-1)/2$, depending on the parity of $N$. $J_{i_0(N-i_0)}$ is hence an integer, the parity of which is dependent on $N\text{mod} 4$. 

In the next step we compare the determinants of $H_e$ and $H_o$. The even part is known to have 0 as an eigenvalue associated with the $N$-partite W state, $\frac{1}{\sqrt{N}}(|100...\rangle+|010...\rangle+...)$. The Hamiltonian commutes with the total angular momentum operator, which in this subspace distinguishes the W state from the rest. Instead of $\text{Det}(H_e)$, we should rather consider the product of nonzero eigenvalues, $\text{Det'}(H_e)=\lim_{\epsilon\rightarrow 0}\text{Det}(H_e+\epsilon)/\epsilon$. The proof of our hypothesis is based on the fact that we expect $\text{Det}'(H_e)$ to be even and at the same time $\text{Det}(H_o)$ to be odd. 

First we consider $N$ odd. We have
\begin{eqnarray}
\text{Det'}(H_e)=\frac{N}{2^{N/2-1}}\prod_{i=1}^{(N-1)/2}J_{i(N-i)},\\
\text{Det}(H_o)=\frac{1}{2^{N/2-1}}\prod_{i=1}^{(N-1)/2}J_{i(N-i)}.
\end{eqnarray}
The second quantity is supposingly odd. However, if we multiply it by $N$, we are expected to get something even, namely $\text{Det'}(H_e)$. Hence a nearest-neighbor Heisenberg chain of odd $N$ cannot be state mirroring.

Next, let us take $N=4n+2$. The determinants are
\begin{eqnarray}
\text{Det'}(H_e)=\frac{2n+1}{2^{2n}}\prod_{i=1}^{2n}J_{i(4n+2-i)},\\
\text{Det}(H_o)=\frac{1}{2^{2n}}\prod_{i=1}^{2n+1}J_{i(4n+2-i)}.
\end{eqnarray}  
To make these two expressions equal, we need to multiply the determinant of $H_o$ by $2n+1$ and the determinant of the reversible part of $H_e$ by an integer $J_{(2n+1)^2}$ (see the comparison of the traces); but again, with multiplying two odd numbers, we expect to get an even one. State mirroring is not possible for $N=4n+2$.

The only exception is, of course, $N=2$, where there is a single gap between the singlet and the triplet.

The argument for the remaining subset of $N=4n$ is a bit less obvious: 
\begin{eqnarray}
\text{Det'}(H_e)=\frac{2n}{2^{2n-1}}\prod_{i=1}^{2n-1}J_{i(4n-i)},\\
\text{Det}(H_o)=\frac{1}{2^{2n-1 }}\prod_{i=1}^{2n}J_{i(4n-i)}.
\end{eqnarray}  
We cannot simply multiply  $\text{Det}(H_o)$ by $2n$ as we would have lost the oddity. Instead, we notice that $\text{Det'}(H_e)$ is supposed to be a product of $2n-1$ even numbers, hence it should involve a factor of $2^{2n-1}$. Then $2n$ can be rewritten as $2^p(2q+1)$, where $p$ and $q$ are integers. One has $p\leq N/2-1$ (the equality holds for $N=4$). Thus we are sure that if the chain satisfies the SPMC, 
\begin{equation}
2^{-p}\text{Det'}(H_e)=\frac{2q+1}{2^{2n-1}}\prod_{i=1}^{2n-1}J_{i(4n-i)}
\end{equation}
is an integer. However, one has
\begin{equation}
(2q+1)\det{Det}(H_o)=2^{-p}J_{4n^2}\det{Det'}(H_e),
\end{equation}
and since $J_{4n^2}$ is even (such is the number of eigenvalues of $H_o$), the left-hand side cannot be odd. 

It is now clear that state mirroring is not a feature of a Heisenberg chain of any number of qubits without local magnetic fields, except for 2. In all cases we have reached the contradiction with the SPMC. 

Finally, let us present a lemma to be used in the summary. The state of one excitation localized at the first site overlaps with all eigenstates of the Hamiltonian.  Was it not the case, there would exist a vector $\vec{v}=(0,c_2,c_3,...)$ which would be an eigenstate of the three-diagonal matrix,
\begin{widetext}
\begin{equation}
\label{eigen}
\frac{1}{2}\left(\begin{array}{cccc} J_{1(N-1)}&-J_{1(N-1)}&0&...\\-J_{1(N-1)}&J_{1(N-1)}+J_{2(N-2)}&-J_{2(N-2)}&...\\0&-J_{2(N-2)}&J_{2(N-2)}+J_{3(N-3)}&...\\...&...&...&...\end{array}\right)\left(\begin{array}{c}0\\c_2\\c_3\\...\end{array}\right)=E\left(\begin{array}{c}0\\c_2\\c_3\\...\end{array}\right).
\end{equation}
\end{widetext}
We have mentioned above that if all coupling constants are non-zero, the only eigenstate with zero energy is the W state, which overlaps with all localized excitations by construction. Thus $E$ is nonzero. The action of the Hamiltonian on $\vec{v}$ produces $(-c_2J_{1(N-1)}/2,[c_2(J_{1(N-1)}+J_{2(N-2)})-c_3J_{3(N-3)}]/2,...)$. Clearly, this requires $c_2=0$ as $J_{1(N-1)}\neq 0$. In the same way we reach $c_i=0$ taking $c_1,...,c_{i-1}=0$. Hence the only vector that can satisfy Eq. (\ref{eigen}) is $\vec{0}=(0,0,0,...,0)$.

In summary, we have shown a fundamental limitation on Heisenberg chains of qubits with nearest neighbor modulated coupling, which prevents them from being mirroring systems, at least in the one excitation subspace \cite{kay}. By the lemma, this makes it impossible for our systems to realize the natural, one-excitation based state transfer, e.g., between the extreme sites, as well as the two sites in the middle for $N$ even, unless one is allowed to introduce local magnetic fields. It remains unclear if it is possible to realize periodic oscillations of the excitation between other pairs of sites. A known special case of ``mirroring'' is the refocusing of the excitation of the middle qubit for $N$ odd for specific choices of couplings. Another interesting question is whether it is possible to naturally transfer a state between the extreme sites of a Heisenberg wire, when one applies the magnetic field only to the first and the last qubits \cite{privcom}. This is well motivated from the experimental point of view, as the ends of the chain are close to heads--devices, which write in and read out the transferred state.

Interestingly, this impossibility emerges even though it was demonstrated that arrays of Heisenberg interacting qubits are suitable for quantum computing \cite{alwayson}. Our result increases the importance of the transfer protocols described above \cite{bgb,danielbose,danielbose1}, and others alike. Which transfer protocol is most feasible will, naturally, depend on the specific system realizing a quantum computer and its relation to the environment.

The Author thanks S. Bose and A. Kay.
This work was supported by the National Research  
Foundation and Ministry of Education, Singapore.


\begin{thebibliography}{99}
\bibitem{jozsa} D. Deutsch and R. Jozsa, {\em Proc. R. Soc. London, Ser. A} {\bf 439}, 553 (1992). 
\bibitem{shor} P .W. Shor, {\em Proceedings of the 35th Annual Symposium Found.
Comp. Sc., Los Alamos} (IEEE Comp. Soc.
Press, Los Alamitos, CA, 1994).
\bibitem{grover} L. Grover, {\em Proceedings of the 28th Annual ACM Symposium
on the Theory of Computing (SOTC)} (ACM Press, New
York, 1996), p. 212; quant-ph/9605043 @ www.arxiv.org.
\bibitem{cluster} R. Raussendorf, D. E. Browne, and H. J. Briegel, {\em Phys. Rev. A} {\bf 68}, 022312 (2003).
\bibitem{teleportation} C. H. Bennett, G. Brassard, C. Cr\'epeau, R. Jozsa, A. Peres, and W. K. Wootters, {\em Phys. Rev. Lett.} {\bf 70}, 1895(1993).
\bibitem{alicki} R. Alicki,  arxiv:0807.2609 @ www.arxiv.org
\bibitem{boseprl91207901} S. Bose, {Phys. Rev. Lett.} {\bf 91}, 207901 (2003).
\bibitem{danielbose} D. Burgarth and S. Bose, {\em Phys. Rev. A} {\bf 71}, 052315 (2005).
\bibitem{chiny} J.-M. Cai, Z.-W. Zhou, G.-C. Guo, {\em Phys. Rev. A} {\bf 74}, 022328 (2006).
\bibitem{kay1} A. Kay, {\em Phys. Rev. Lett.} {\bf 98}, 010501 (2007).
\bibitem{exp3} D. Tsomokos, M. Hartmann, S. Huelga, and M. B. Plenio, {\em New. J. Phys.} {\bf 9}, 79 (2007).
\bibitem{thermal} M. Wie\'sniak, arxiv:0711.2357 @ www.arxiv.org
\bibitem{distill} C. H. Bennett, G. Brassard, S. Popescu, B. Schumacher, J.A. Smolin, and W. K. Wootters, {\em Phys. Rev. Lett.} {\bf 76}, 722 (1996).

\bibitem{christ} M. Christandl, N. Datta, A. Ekert, and A. J. Landahl, {\em Phys. Rev. Lett.}	{\bf 92}, 187902 (2004).
\bibitem{niko1} G. M. Nikolopoulos, D. Petrosyan, and P. Lambropoulos, {\em Eurphys. Lett.} {\bf 65}, 297 (2004).
\bibitem{niko2} G. M. Nikolopoulos, D. Petrosyan, and P. Lambropoulos, {\em Jour. Phys.:Cond. Matt.} {\bf 16}, 4991 (2004).
\bibitem{Li}  Y. Li, T. Shi, B. Chen, Z. Song, and C. P. Sun, {\em Phys. Rev. A} {\bf 71} 022301 (2005).
\bibitem{Shi} T. Shi, Y. Li, Z. Song, and C. P. Sun, {\em Phys. Rev. A} {\bf 71}, 032309 (2005). 
\bibitem{christ1} M. Christandl, N. Datta, T. C. Dorlas, A. Ekert, A. Kay, and A. J. Landahl, {\em Phys. Rev. A}	{\bf 71}, 032312 (2005).
\bibitem{danielbose1} D. Burgarth and S. Bose, {\em New Jour. Phys.} {\bf 7}, 135 (2005).
\bibitem{bgb} D. Burgarth, V. Giovannetti, and S. Bose, {\em Phys. Rev. A} {\bf 75}, 062327 (2007).
\bibitem{dago} T. Oshima, A. Ekert, D. K. L. Oi, D. Kaszlikowski, and L.C. Kwek, quant-ph/0702019 @ www.arxiv.org.
\bibitem{paternostro} C. DiFranco, M. Paternostro, and M. S. Kim, arxiv:0805.4365 @ www.arxiv.org.
\bibitem{mirror} C. Albanese, M. Christandl, N. Datta, and A. Ekert, {\em Phys. Rev. Lett.} {\bf 93}, 230502 (2004).
\bibitem{cavities} D. G. Angelakis, M. F. Santos, and S. Bose, {\em Phys. Rev. A} {\bf 76}, 031805(R) (2007).
\bibitem{nadeslane} M. J. Hartmann, F. G. S. L. Brand\~{a}o, and M. Plenio, {\em Phys. Rev. Lett.} {\bf 99}, 160501(2007).
\bibitem{dots1} D. Loss and D. P. DiVincenzo, {\em Phys. Rev. A} {\bf 57}, 120 (1998).
\bibitem{dots2} B. E. Kane, {\em Nature} {\bf 393}, 133 (1998),
\bibitem{dots3} R. Vrijen {\em et al.}, {\em Phys. Rev. A}, {\bf 62}, 012306 (2000).
\bibitem{katsura} S. Katsura, {\em Phys. Rev} {\bf 127}, 1508 (1962).
\bibitem{privcom} D. Burgarth (private communication).
\bibitem{alwayson} S. C. Benjamin and S. Bose, {\em Phys. Rev. Lett.} {\bf. 90}, 247901 (2003). 
\bibitem{kay} For the proof of no mirroring in square $xx$ lattices, see A. Kay, quant-ph/0702088 @ www.arxiv.org.
\end{thebibliography}
\end{document}